%% file: RJwrapper.tex
\documentclass[a4paper]{report}
\usepackage[utf8]{inputenc}
\usepackage[T1]{fontenc}
\usepackage{RJournal}
\usepackage{amsmath,amssymb,array}
\usepackage{booktabs}

\begin{document}

\sectionhead{Contributed research article}
\volume{XX}
\volnumber{YY}
\year{20ZZ}
\month{AAAA}

\begin{article}
  \input{tang-li}

\end{article}

\end{document}

%% file: tang-li.tex
\title{lfda: An R Package for Local Fisher Discriminant Analysis and Visualization}
\author{by Yuan Tang and Wenxuan Li}

\maketitle

\abstract{
Local Fisher discriminant analysis is a localized variant of Fisher discriminant analysis and it is popular for supervised dimensionality reduction method. \pkg{lfda} is an R package for performing local Fisher discriminant analysis, including its variants such as kernel local Fisher discriminant analysis and semi-supervised local Fisher discriminant analysis.  It also provides  visualization functions to easily visualize the dimension reduction results by using either \pkg{rgl} for 3D visualization or \pkg{ggfortify} for 2D visualization in \pkg{ggplot2} style. 
}


\section{Introduction}

Fisher discriminant analysis \citep{fdapaper} is a popular choice to reduce the dimensionality of the original data set. It maximizes between-class scatter and minimizes within-class scatter. It works really well in practice but lacks some considerations for multimodality. Multimodality exists within many applications, such as disease diagnosis, where there may be multiple causes for a particular disease. In this situation, Fisher discriminant analysis cannot capture the multimodal characteristics of the clusters. To deal with multimodality, local-preserving projection \citep{lpppaper} preserves the local structure of the data in that it keeps nearby data pairs in the original data space close in the embedding space. As a result, multimodal data could be embedded and its local structure will not be lost.

Later on, a new dimensionality reduction method called local Fisher discriminant analysis \citep{lfdapaper} was proposed to combine both advantages of Fisher discriminant analysis and those of local-preserving projection in a way that between-class separability is maximized while within-class separability is minimized and its local structure is preserved. Furthermore, with the help of kernel trick, local Fisher discriminant analysis can also be extended to deal with non-linear dimensionality reduction situations.

Principal components analysis \citep{pcapaper} is another popular choice for performing dimension reduction. However, in practice sometimes principal components analysis generates bad principal components that cannot explain a great amount of variance in the original data set. For example, if the original data set has six dimensions and we reduce the dimension into three using PCA, the resulted three principal components might not capture some important features and variance in the original data features. Using this result from PCA sometimes lead to very bad accuracy for machine learning models due to the poor pre-processing that losses a lot of the essential information in the original data features. On the other hand, metric learning methods such as LFDA in particular can, surprisingly, enhance the distinctive characteristics of the original data set and pull data points that have similar characteristics close to each other. However, both PCA and LFDA have their own shortcomings and a combined approach called semi-supervised local Fisher discriminant analysis \citep{semilfdapaper} mix the supervised and unsupervised approaches to provide a more stable result. We will briefly discuss this in the following sections. 

In this paper, we will introduce an R package called \CRANpkg{lfda} \citep{lfda} . It's is an R package for performing local Fisher discriminant analysis as well as its variants such as kernel local Fisher discriminant analysis and semi-supervised local Fisher discriminant analysis.  It also provides  visualization functions to easily visualize the dimension reduction results by using either \CRANpkg{rgl} \citep{rgl} for 3D visualization or using \CRANpkg{ggfortify} \citep{ggfortify} for 2D visualization in \CRANpkg{ggplot2} \citep{ggplot2} style. 

\section{Theoretical background}

\subsection{Fisher linear discriminant analysis (FDA)}

Fisher linear discriminant analysis is a popular method used to find a linear combination of features that characterizes or separates two or more classes of objects and events. 

Let $S^{(w)}$ and $S^{(b)}$ be the \textit{within-class scatter matrix} and the \textit{between-class scatter matrix} defined by the following:

$$S^{(w)}=\sum_{i=1}^{l}\sum_{j:y_{j}=i} (x_{j}-\mu_{i})(x_{j}-\mu_{i})^{T},$$
and
$$S^{(b)}=\sum_{i=1}^{l}n_{i} (\mu_{i}-\mu)(\mu_{i}-\mu)^{T},$$

where $\mu_{i}$ is the mean of samples in the class $i$ and $\mu$ is the mean of all samples, such that:
$$\mu_{i}=\frac{1}{n_i}\sum_{j:y_{j}=i}x_{j},$$
$$\mu=\frac{1}{n}\sum_{i=1}^{n}x_{i}.$$

The FDA transformation matrix, denoted as $T_{FDA}$, is defined as follows:
$$T_{FDA}=argmax_{T\in \mathbb{R}^{d\times m}}((T^{T}S^{(w)}T)^{-1}T^{T}S^{(b)}T).$$

In other words, $T$ is found so that between-class scatter is maximized and within-class scatter is minimized. In order to optimize this, $T_{FDA}$ can be found by the following:
$$T_{FDA}=(\phi_{1}|\phi_{2}|\cdots|\phi_{m}),$$
where $\{\phi_{i}\} _{i=1}^{d}$ are the generalized eigenvectors associated to the generalized eigenvalues $\lambda_{1}\geq \lambda_{2} \geq \cdots \geq \lambda_{d}$ of the generalized eigenvalue problem: $$S^{(b)}\phi = \lambda S^{(w)}\phi.$$

\subsection{Locality-preserving projection (LPP)}

Locality preserving projections are linear projective maps that arise by solving a variational problem that optimally preserves the neighborhood structure of the data set.

Let $A$ be the \textit{affinity matrix} where its $(i,j)$-th element is the affinity between two points $x_{i}$ and $x_{j}$. The elements in $A$ ia larger if two points are close to each other. We define $A_{i,j}=1$ if $x_{j}$ is the $k$-nearest neighbor of $x_{i}$ or vice versa while $A_{i,j}=0$ otherwise. The \textit{LPP transformation matrix} can then be defined as the following:
$$T_{LPP}=argmin_{T\in \mathbb{R}^{d\times m}}\frac{1}{2}\sum_{i,j=1}^{n}A_{i,j}||T^{T}x_{i}-T^{T}x_{j}||^{2}$$
subject to $$T^{T}XDX^{T}T=I,$$
where $D$ is the diagonal matrix with $i$-th diagonal element being $$D_{i,i}=\sum_{j=1}^{n}A_{i,j}.$$

$T_{LPP}$ is found such that nearby data pairs in the original data space are kept close in the embedding space. 
$T_{LPP}$ is given by $$T_{LPP}=(\phi_{d-m+1}|\phi_{d-m+2}|\cdots|\phi_{d}),$$
where $\{\phi_{i}\}_{i=1}^{d}$ are the eigenvectors associated to the eigenvalues  $\lambda_{1}\geq \lambda_{2} \geq \cdots \geq \lambda_{d}$ of the following eigenvalue problem $$XLX^{T}\phi=\lambda XDX^{T}\phi,$$
where $$L=D-A.$$

\subsection{Local Fisher discriminant analysis (LFDA)}

We combine the concepts for both FDA and LPP to define local Fisher discriminant analysis. A \textit{local} within-class scatter matrix $$S^{(w)}=\frac{1}{2}\sum_{i,j=1}^{n}W_{i,j}^{(w)}(x_{i}-x_{j})(x_{i}-x_{j})^{T},$$
 $$S^{(b)}=\frac{1}{2}\sum_{i,j=1}^{n}W_{i,j}^{(b)}(x_{i}-x_{j})(x_{i}-x_{j})^{T},$$
\\
where 

$$
W_{i,j}^{(w)}=
\left\{
	\begin{array}{ll}
		A_{i,j}/n_{l}  & \mbox{if } y_{i}=y_{j}=l, \\
		0 & \mbox{if } y_{i}\neq y_{j},
	\end{array}
\right.
$$

$$
W_{i,j}^{(b)}=
\left\{
	\begin{array}{ll}
		A_{i,j}(1/n-1/n_{l})  & \mbox{if } y_{i}=y_{j}=l, \\
		1/n & \mbox{if } y_{i}\neq y_{j}.
	\end{array}
\right.
$$

Intuitively, we use the above equations to weight the values for each pair of samples that belong to the same class. Namely, for sample pairs that are far apart from each other in the same class, we give less weight/influence on the two local scatter matrices.

The LFDA transformation matrix $T_{LFDA}$ is then defined as the following:
$$T_{LFDA}=argmax_{T\in \mathbb{R}^{d \times r}}\left[ tr((T^{T}S^{(w)}t)^{-1}T^{T}S^{(b)}T)\right]$$

The transformation matrix $T_{LFDA}$ is found so that data pairs that are near each other in the same class become closer and data pairs in different classes become far away from each other. Note that data pairs in the same class that are far away from each other do not come closer.

\subsection{Kernel local Fisher discriminant analysis (KLFDA)}
LFDA can be \textit{non-linearized} by the \textit{kernel trick}, which enables the algorithm to operate in a high-dimensional feature space but avoids computing the coordinates of the data in that space. It works by computing the inner products between all pairs of data in the feature space. Kernel trick also greatly reduces the costs of computation of the coordinates. This then becomes kernel local Fisher discriminant analysis \citep{lfdapaper}. \\

\noindent One particular choice of kernel function is the Gaussian kernel defined as the following:
$$K(x,x')=exp(-\frac{||x-x'||^{2}}{2\sigma ^{2}}),$$
with $\sigma>0.$ $K$ is now the \textit{kernel matrix} where its $(i,j)$-th element is given by $$K_{i,j}=<\phi(x_{i}),\phi(x_{j})>=K(x_{i},x_{j}),$$
which we then use to perform the regular LFDA. 
\subsection{Semi-supervised local Fisher discriminant analysis (SELF)}

The performance of supervised dimensionality reduction methods, such as LFDA, tend to be degraded when 
only a small number of labeled samples are available. Thus, the supervised methods overfit embedding spaces to the labeled samples. On the other hand, unsupervised dimensionality reduction methods, such as principal components analysis, preserve the global data structure really well. Semi-supervised local Fisher discriminant analysis smoothly bridges LFDA and PCA  so that our reliance on the global structure of unlabeled samples and information brought by labeled samples can be controlled. 

\section{Software architecture}

This package fully utilizes S3 generic functions (\url{http://adv-r.had.co.nz/S3.html}) to recognize \code{lfda} objects obtained from \code{lfda()}, \code{klfda()}, or \code{self()}. 

Each \code{lfda} object is a list consists of the following items:

\begin{itemize}
	\item \code{T}: the trained transformation matrix
	\item \code{Z}: the transformed original data with reduced dimensionality
\end{itemize}

The following methods can be applied to \code{lfda} objects:

\begin{itemize}
  \item \code{predict()}: generate predictions on new data
  \item \code{print()}: print out summary statistics of the \code{lfda} object, including the trained transformation matrix and the original data after applying the transformation matrix
  \item \code{plot()}: generate 3D visualization with the help from \pkg{rgl} package
\end{itemize}

The integration with \pkg{ggfortify} recognizes the \code{lfda} class and then plots the transformed data that's similar to its PCA visualizations. We will illustrate this by providing some examples in the next section. 

\section{Illustrations}

In the following sub-sections, we will perform LFDA and its variants on the famous (Fisher's or Anderson's) iris data set that gives the measurements in centimeters of the variables sepal length and width and petal length and width, respectively, for 50 flowers from each of 3 species of iris, which are Iris setosa, versicolor, and virginica.

\subsection{Local Fisher discriminant analysis}

\begin{figure}[htbp]
  \centering
  \includegraphics[width=145mm,scale=0.8]{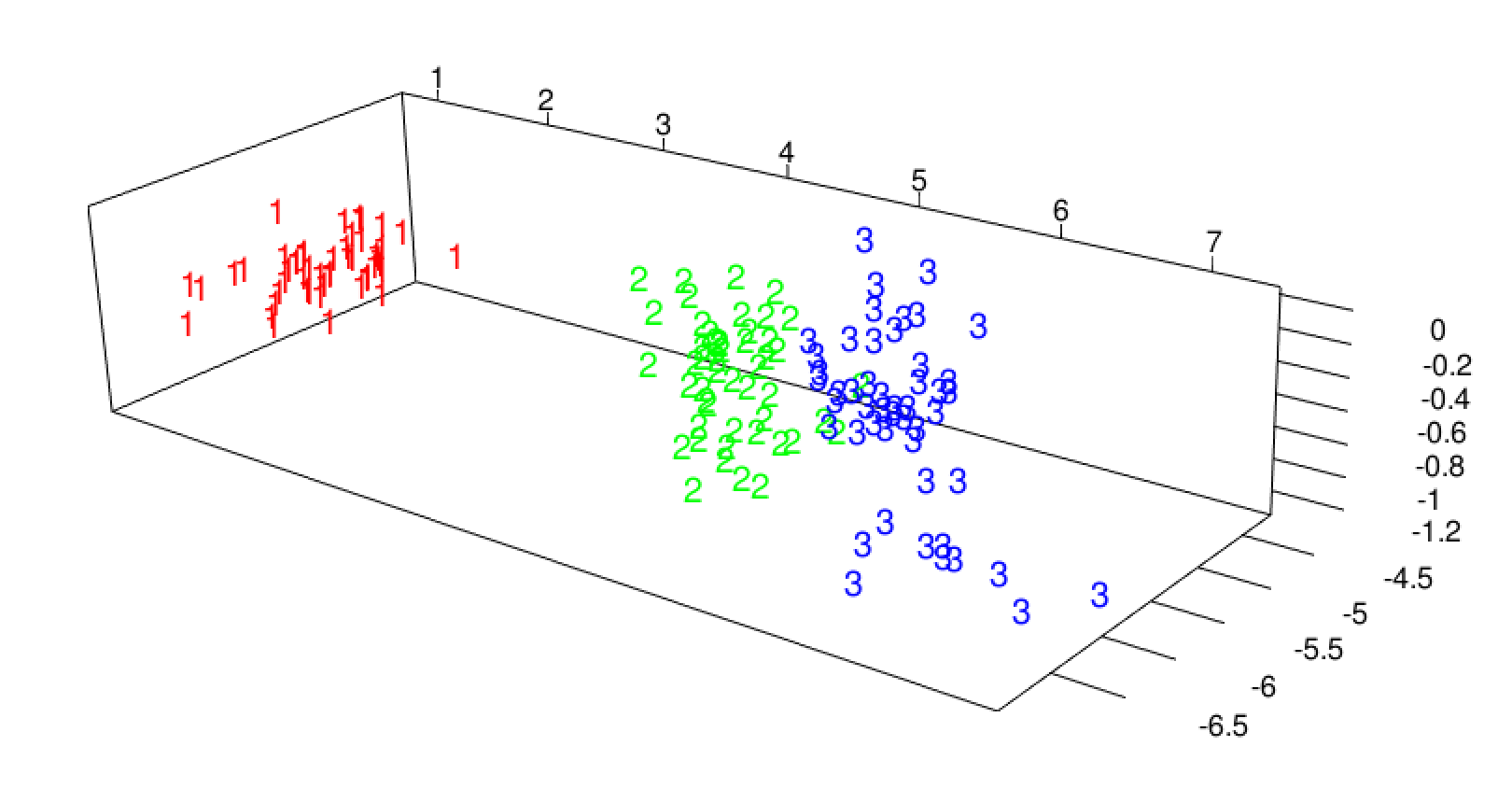}
  \caption{Local Fisher discriminant analysis 3D visualization. }
  \label{figure:lfda3d}
\end{figure}

We pass the original iris samples and its labels into \code{lfda()} as \code{x} and \code{y}, respectively, along with \code{r} that represents the desired dimensionality to reduce the original data into and the \code{metric} that switches the type of returned metric. The available types of metric are weighted, orthonormalized, and plain. Once We trained a local Fisher discriminant analysis model, we can then call \code{predict()} to transform new data set using the trained distance matrix. We can then visualize the LFDA model in 3D by calling \code{plot()}. See Figure~\ref{figure:lfda3d} for an example visualization. 

\begin{example}
library(lfda)
data(iris)
x <- iris[, -5]
y <- iris[, 5]
r <- 3
model <- lfda(x, y, r, metric = "plain")
transformedData <- predict(model, iris[, -5])
plot(x = model, labels = iris[, 5])
\end{example}

\subsection{Kernel local Fisher discriminant analysis}

\begin{figure}[htbp]
  \centering
  \includegraphics[width=145mm,scale=0.8]{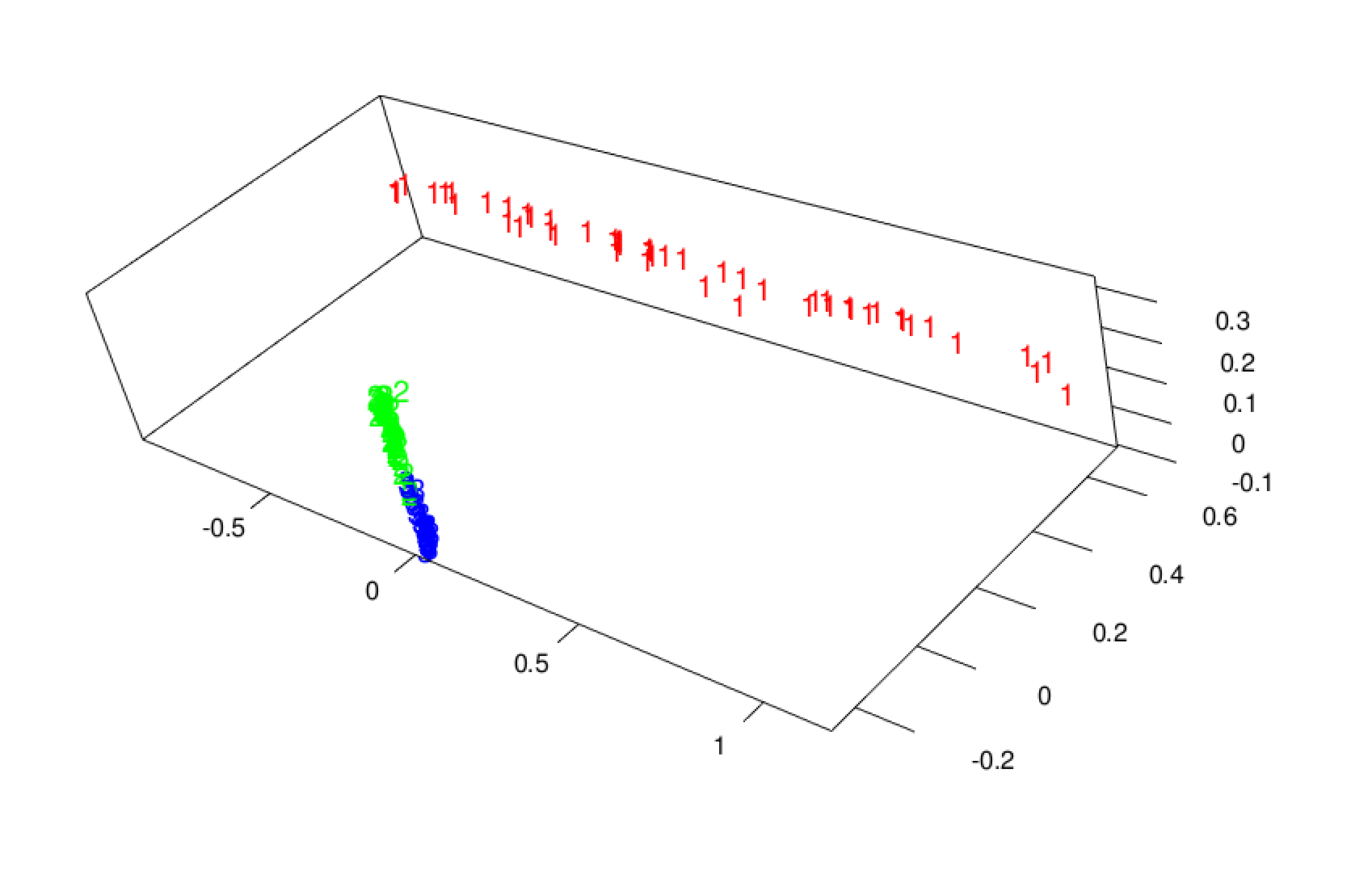}
  \caption{Kernel local Fisher discriminant analysis 3D visualization. }
  \label{figure:klfda3d}
\end{figure}

The following code performs kernel local Fisher discriminant analysis. We first apply \code{kmatrixGauss()} provided in \pkg{lfda} to compute Gaussian kernel matrix of the original data set, which maps the original data space to non-linear and higher dimensions. We can then pass this kernel matrix into \code{klfda()} to perform KLFDA. Similarly, we can plot the transformed data in 3D. See Figure~\ref{figure:klfda3d} for an example visualization. 

\begin{example}
library(lfda)
data(iris)
k <- kmatrixGauss(iris[, -5])
y <- iris[, 5]
r <- 3
model <- klfda(k, y, r, metric = "plain")
plot(x = model, labels = iris[, 5])
\end{example}

Note that for iris data set, the relationships between different classes are linear. Kernel Local Fisher Discriminant Analysis is only aimed for capturing non-linear relationships, especially when it comes to many different classes. In this case, visualization of iris data set is poor because KLFDA is too strong for capturing linear relationships. If using KLFDA for this kind of data, later when it comes to classification or clustering tasks, the model would very likely overfit the transformed data set.

\subsection{Semi-supervised local Fisher discriminant analysis}

We can perform semi-supervised local Fisher discriminant analysis in a way similar to LFDA. In addition, we have three more parameters to use:

\begin{itemize}
	\item \code{beta}: the degree of semi-supervisedness with value between 0 and 1 where 0 represents totally supervised (discard all unlabeled samples) and 1 represents totally unsupervised (discard all labeled samples).
	\item \code{kNN}: the parameter used in local scaling method
	\item \code{minObsPerLabel}: the minimum number observations required for each different label
\end{itemize}

\begin{figure}[htbp]
  \centering
  \includegraphics[width=145mm,scale=0.8]{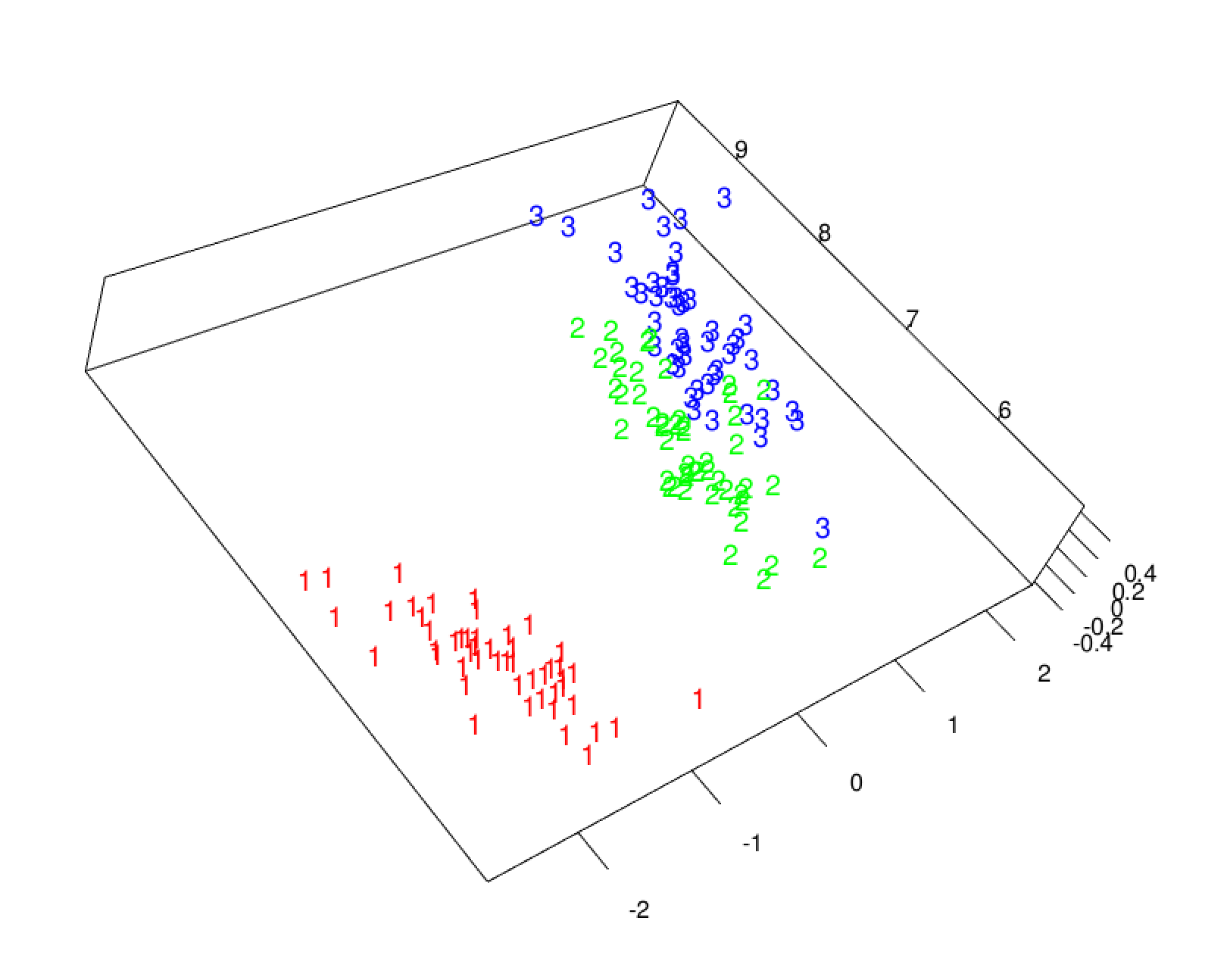}
  \caption{Semi-supervised local Fisher discriminant analysis 3D visualization. }
  \label{figure:self3d_beta01}
\end{figure}

Here we discard $10\%$ of the labels and perform semi-supervised local Fisher discriminant analysis that smoothly bridges supervised LFDA and unsupervised principal component analysis, by which a natural regularization effect can be obtained when only a small number of labeled samples are available. See Figure~\ref{figure:self3d_beta01} for an example 3D visualization. 

\begin{example}
library(lfda)
data("iris")
x <- iris[, -5]
y <- iris[, 5]
r <- 3
model <- self(x, y, beta = 0.1, r = 3, metric = "plain")
plot(model, iris[, 5])
\end{example}

\subsection{Visualize using \pkg{ggfortify}}

\begin{figure}[htbp]
  \centering
  \includegraphics[width=145mm,height=100mm,scale=0.8]{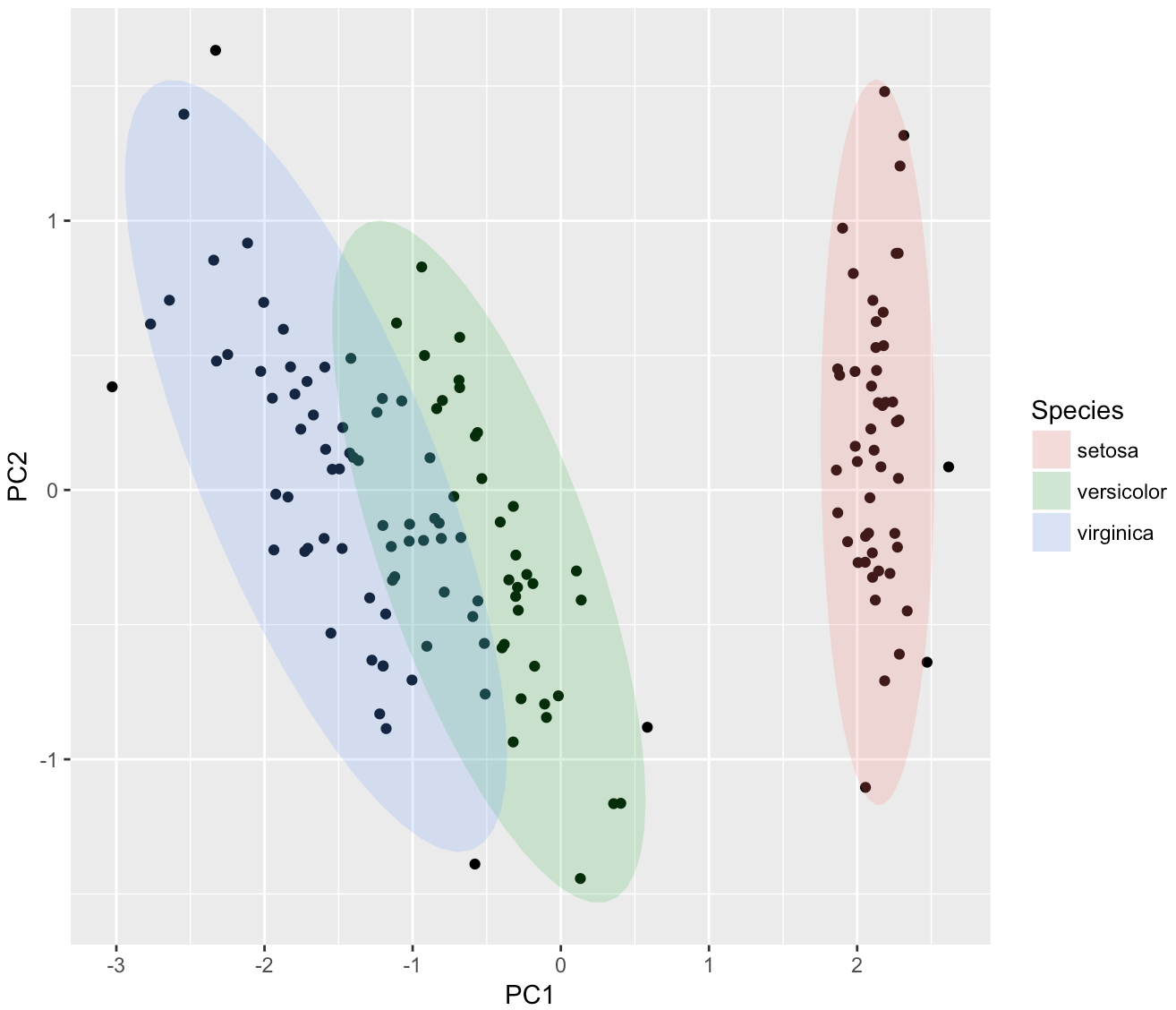}
  \caption{Semi-supervised local Fisher discriminant analysis visualization using \pkg{ggfortify}.}
  \label{figure:ggfortify-lfda}
\end{figure}

\pkg{lfda} is integrated with \pkg{ggfortify} so you can use \code{autoplot()} to automatically plot \code{lfda} objects in 2D with \pkg{ggplo2} style. We can specify \code{frame = TRUE} to draw the frame boundary as well as specifying  \code{frame.type} for shape of the frame and \code{frame.colour} for the column to be used to color the frame. More options can be found in \pkg{ggfortify} documentation as well as the \pkg{ggfortify} paper on the R Journal \citep{ggfortifypaper}. All options available in clustering and PCA in \pkg{ggfortify} can be applied to \code{lfda} objects. See Figure~\ref{figure:ggfortify-lfda} for an example. 

\begin{example}
library(lfda)
library(ggfortify)
data(iris)
model <- self(iris[, -5], iris[, 5], beta = 0.1, r = 3, metric = "plain")
autoplot(model, data = iris, frame = TRUE, frame.colour = 'Species', frame.type = 'norm')
\end{example}

\section{Future development}

\noindent In the future, users will be able to use methods exported from \pkg{lfda} inside \CRANpkg{dml} \citep{dml}, which is a collection of distance metric learning algorithm implementations so users can compare results from different algorithms very easily. Any suggestions collected from StackOverflow, Github issues, etc, as well as contributions are welcomed. The package is hosted on Github at \url{https://github.com/terrytangyuan/lfda}. 

\section{Summary}

\pkg{lfda} provides a simple interface to conduct local Fisher discriminant analysis and its various variants, such as kernel local Fisher discriminant analysis and semi-supervised local Fisher discriminant analysis. It also provides  visualization functions to easily visualize the dimension reduction results by using either \pkg{rgl} for 3D visualization or  \pkg{ggfortify} for 2D visualization in \pkg{ggplot2} style. 

\section{Acknowledgement}

We sincerely thank all developers for their efforts behind the packages that \pkg{lfda} depend on, namely, \CRANpkg{plyr} \citep{plyr}, \CRANpkg{rARPACK} \citep{rARPACK}, and \CRANpkg{rgl} \citep{rgl}. We thank Zachary Deane-Mayer for helping setting up the package structure and continuous integration using Travis CI and thank Nan Xiao for his initial work-in-progress code.


\bibliography{tang-li}

\address{Yuan Tang\\
  Uptake Technologies, Inc. \\
  600 West Chicago Ave, Chicago, IL 60654\\
  United States\\}
\email{terrytangyuan@gmail.com}

\address{Wenxuan Li\\
  Department of Agricultural Economics, Purdue University\\
  403 W State Street, West Lafayette, IN, 47907\\
  United States\\}
\email{wenxuan.tess@gmail.com}